\documentclass[
    amsmath,
    amssymb,
    reprint,
    aps,
    prx,
    superscriptaddress,
    longbibliography
]{revtex4-1}

\usepackage[utf8]{inputenc}
\usepackage[T1]{fontenc}
\usepackage{physics}
\usepackage{nicefrac}
\usepackage{bbm, bm}
\usepackage{makecell}
\usepackage{multirow}
\usepackage{booktabs}
\usepackage{overpic}
\usepackage{tikz}
\usepackage{pgfplots}
\pgfplotsset{compat=1.18}
\usepackage[dvipsnames]{xcolor}

\usepackage[breaklinks=true, colorlinks]{hyperref}
\hypersetup{linkcolor=blue, citecolor=blue, urlcolor=MidnightBlue}
\usepackage{cleveref}

\begin{document}

\title{Efficient local counterdiabatic driving}

\author{Alev Orfi}
\affiliation{Center for Quantum Phenomena, Department of Physics, New York University, 726 Broadway, New York, New York 10003, USA}
\affiliation{Center for Computational Quantum Physics, Flatiron Institute, 162 5th Avenue, New York, NY 10010, USA}

\author{Dries Sels}
\affiliation{Center for Computational Quantum Physics, Flatiron Institute, 162 5th Avenue, New York, NY 10010, USA}
\affiliation{Department of Physics, Boston University, 590 Commonwealth Ave., Boston, Massachusetts 02215, USA}

\date{\today}

\begin{abstract}
Counterdiabatic (CD) driving removes the need for slow driving in adiabatic state preparation by suppressing unwanted transitions, but computing and implementing CD drives remain challenging in many-body quantum systems. 
Several recent schemes construct universal approximations to these drives through Krylov expansions, although rapid operator growth typically leads to slow convergence.
Here, we introduce an improved \emph{local} expansion for geometrically local Hamiltonians and prove that it reduces the sufficient expansion order from nearly linear to polylogarithmic in system size, provided the state of interest is separated by a finite spectral gap.
The resulting local adiabatic gauge potential further provides a systematic construction for transporting local observables along gapped Hamiltonian paths with resources independent of system size.
\end{abstract}

\maketitle

\emph{Introduction--}~
Counterdiabatic (CD) driving enables rapid adiabatic state preparation by suppressing unwanted transitions. Its implementation, however, requires constructing the adiabatic gauge potential (AGP)~\cite{delcampo13,sels17,Odelin19}, which is generally computationally intractable and highly non-local in chaotic many-body systems~\cite{pandey20}. This challenge has motivated a variety of approximate approaches, including mean-field~\cite{hatomura17,hayasaka2023generalmethodconstructmean} and variational~\cite{saberi14,sels17} driving schemes. 
Local variational approaches are particularly attractive, as they do not require knowledge of the instantaneous eigenstates and allow systematic improvement through increasingly expressive operator ansatzes~\cite{sels17}. Their locality also facilitates implementation on digital and analog quantum devices~\cite{Claeys19,hegade22}. A particularly convenient choice of ansatz is provided by nested-commutator expansions, whose operators belong to a Krylov subspace of the Liouvillian~\cite{Claeys19,Takahashi24,hatomura2026universaldigitizedcounterdiabaticdriving}. However, achieving high accuracy can require high commutator orders, leading to increasingly large and computationally demanding operator expansions.
Recent polynomial approximations provide rigorous convergence guarantees, but their required expansion order increases with system size due to the extensive many-body bandwidth, $B\sim N$, even when the spectral gap remains finite~\cite{finzgar2025counterdiabatic}.
As discussed in detail in Ref.~\cite{morawetz25} the convergence of this Krylov scheme is controlled by the asymptotic high frequency tails of the response functions, which control the operator growth in Krylov space~\cite{parker19}. In typical many-body systems, operators spread exponentially fast in Krylov space, leading to logarithmically slow convergence of the AGP in the thermodynamic limit~\cite{finzgar2025counterdiabatic}. Here we show how one can largely eliminate the effect of operator growth, resulting in a significantly faster converging scheme while retaining a structure that makes it easy to implement on quantum devices. 

In condensed matter physics, approximate counterdiabatic driving has found applications as a scheme to perform quasi-adiabatic continuation of many-body ground states~\cite{hastings05,bachman12}. The latter provides a rigorous framework for defining and classifying gapped phases of matter. The covariance matrix of the AGPs defines the quantum geometric tensor, whose real part is the quantum metric~\cite{provost80,KOLODRUBETZ20171}, which indicates the presence of a phase transition, and reveals its universal properties, without prior knowledge of any order parameter~\cite{polkovnikov05,venuti07,wang15}.

Here, we exploit the geometric locality of the Hamiltonian to construct an approximate local AGP with rigorous convergence guarantees.
We demonstrate its advantages for counterdiabatic driving and establish a scheme for transporting local observables. 
The latter enables the classical computation of local expectation values within gapped phases with resources independent of system size and quasipolynomial in the inverse target accuracy.

\emph{Local AGP approximation--}~
Consider a smooth family of Hamiltonians $H(\lambda)$, with $\lambda\in[0,1]$ and instantaneous eigenstates $H(\lambda)|\phi_n(\lambda)\rangle=E_n(\lambda)|\phi_n(\lambda)\rangle$. Assume $\partial_\lambda H=\sum_j O_j$, comprising of $\mathcal O(N)$ geometrically local terms. Eigenstates can be adiabatically transported by the AGP, as
\begin{equation}
    i \partial_\lambda |\phi_n(\lambda)\rangle=A_\lambda|\phi_n(\lambda)\rangle,
\end{equation}
where the AGP has matrix elements
\begin{equation}
    \langle\phi_m|A_\lambda|\phi_n\rangle
    =-i\frac{\langle\phi_m|\partial_\lambda H|\phi_n\rangle}
    {\omega_{mn}},\qquad \omega_{mn}\neq0,
\end{equation}
with $\omega_{mn}=E_m-E_n$. Introducing the Liouvillian $\mathcal L_\lambda(O)=[H(\lambda),O]$, the AGP can be written as
\begin{equation}
    A_\lambda
    =-ig(\mathcal L_\lambda)\partial_\lambda H
    =-i\sum_jg(\mathcal L_\lambda)O_j,
\end{equation}
where $g(\omega)=1/\omega$ for $\omega\neq0$ and $g(0)=0$ fixes the gauge. 
We assume a nonzero spectral spacing $\omega_{\rm min}$ and define the maximum bandwidth $B$ along the path as
\begin{equation}
    \begin{aligned}
    \omega_{\rm min}
    &=\min_{\lambda\in[0,1]}
    \min_{\omega_{mn}(\lambda)\neq0}
    |\omega_{mn}(\lambda)|>0,\\
    B&=\max_{\lambda\in[0,1]}
    \max_{m,n}|\omega_{mn}(\lambda)|.
    \end{aligned}
\end{equation}
Thus, approximating the AGP reduces to approximating $g(\omega)=1/\omega$ over $[-B,-\omega_{\rm min}]\cup[\omega_{\rm min},B]$, with $g(0)=0$. For transport of a specific eigenstate, such as the ground state, only a subset of these frequencies is relevant, as discussed in the next section. 
Existing polynomial approaches approximate the AGP through nested-commutator expansions whose required order increases with the ratio $B/\omega_{\rm min}$ and therefore typically with system size~\cite{Claeys19,Takahashi24,finzgar2025counterdiabatic}.

Instead of directly constructing a polynomial approximation to the AGP, we first introduce a finite-time filter ~\cite{hastings05,wan2020fast} that allows the locality of the Hamiltonian to be exploited. Consider the filter
\begin{equation}
    f_T(\omega)
    =\operatorname{sinc}^M\left(\frac{T\omega}{M}\right)
    =\int_{-T}^T dt\,p_T(t)e^{i\omega t},
\end{equation}
where $p_T(t)$ is the $M$-fold convolution of uniform distributions on $[-T/M,T/M]$. Choosing $T=2M/\omega_{\rm min}$ ensures $|f_T(\omega)|\leq 2^{-M}$ for all $|\omega|\geq\omega_{\rm min}$. This provides a finite-time approximation to $g(\omega)$ of the form
\begin{equation}
    g_T(\omega)=\frac{1-f_T(\omega)}{\omega},
    \qquad g_T(0)=0,
\end{equation}
with error
\begin{equation}
    \epsilon_g
    =\sup_{\omega_{\rm min}\leq|\omega|\leq B}
    \left|g_T(\omega)-\frac1\omega\right|
    \leq\frac{2^{-M}}{\omega_{\rm min}}.
\end{equation}
Thus, the spectral approximation converges exponentially with $M$, independently of the bandwidth $B$.

Achieving a relative spectral error $\omega_{\rm min}\epsilon_g\leq\epsilon$ requires only $M=\lceil\log_2(1/\epsilon)\rceil$ convolutions, corresponding to a maximum evolution time
\begin{equation}
    T=\mathcal O\left(
    \frac1{\omega_{\rm min}}\log\frac1\epsilon
    \right).
    \label{eq:Timebound}
\end{equation}
Crucially, the finite support of $p_T(t)$ allows the approximate AGP to be expressed in terms of finite-time Heisenberg evolution. Defining $O_j(t)=e^{iHt}O_je^{-iHt}$ gives
\begin{equation}\label{eq:finite_time_agp}
    A_{\lambda,T}
    =\sum_j\int_0^Tdt\,p_T(t)
    \int_0^t ds\,[O_j(-s)-O_j(s)].
\end{equation}
Since $|s|\leq T$, each contribution can be localized by restricting the Hamiltonian to a region of radius $R$ around the support of $O_j$.
Defining $O_{j,R}(t)=e^{iH_{R,j}t}O_je^{-iH_{R,j}t}$, Lieb--Robinson bounds~\cite{kliesch2014lieb} imply
\begin{equation}
    \|O_j(t)-O_{j,R}(t)\|_{\rm op}
    \leq C e^{-\mu(R-v|t|)},
\end{equation}
where $v$ is a Lieb--Robinson velocity and $C,\mu>0$ are constants independent of system size. Each local contribution to the finite-time AGP can be approximated by
\begin{equation}\label{eq:localized_agp}
    A_{j,T,R}
    =\int_0^Tdt\,p_T(t)
    \int_0^t ds\,[O_{j,R}(-s)-O_{j,R}(s)],
\end{equation}
with operator-norm error
\begin{equation}
    \|A_{j,T}-A_{j,T,R}\|_{\rm op}
    \leq CTe^{-\mu(R-vT)}.
\end{equation}
Thus, an error $\epsilon_R$ per contribution is achieved with
\begin{equation}
    R\geq vT+\mu^{-1}\log(CT/\epsilon_R).
\end{equation}
Combining this with the finite-time bound in Eq.~\eqref{eq:Timebound} gives,
\begin{equation}
    R=\mathcal O\left(
    \frac{1}{\omega_{\rm min}}\log\frac1\epsilon
    \right),
\end{equation}
for comparable filtering and localization tolerances.
Each contribution can therefore be approximated independently using a restricted Hamiltonian with bandwidth 
\begin{equation}
    B_R\leq2\|H_{R,j} \|= \mathcal O(R^d)
    =\mathcal O\left(\frac{1}{\omega_{\rm min}^d}\log^d\frac1\epsilon\right).
\end{equation}
For a minimum spectral spacing bounded below independently of system size, the restricted bandwidth is independent of the total system size. This reduction in spatial support and bandwidth can substantially lower the computational resources required by existing polynomial and variational AGP approximations.

\begin{figure}[t]
    \centering
    \includegraphics[width=\columnwidth]{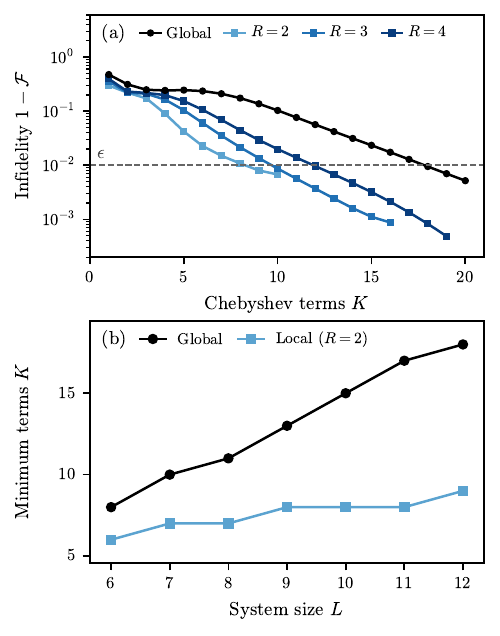}
    \caption{
    Localized Chebyshev approximations for ground-state transport in the mixed-field Ising chain.
    (a) Final infidelity versus the number of odd Chebyshev terms $K$ for $L=12$, comparing global and localized expansions of radius $R$.
    (b) Minimum $K$ required to achieve $1-\mathcal F\leq10^{-2}$ as a function of system size.
    All approximations are constructed directly for $1/\omega$, without finite-time filtering.}
    \label{fig:figure1}
\end{figure}

\emph{Ground-state transport --}~ 
We now consider the preparation of a many-body ground state by transporting an initial ground state $|\phi_0(0)\rangle$ along the Hamiltonian path. Exact counterdiabatic driving keeps the system in the instantaneous ground state, independently of the driving rate, and is generated by
\begin{equation}
    H_{\rm CD}(t)
    =H(\lambda(t))+\dot\lambda(t)A_\lambda.
\end{equation}
Replacing the exact AGP $A_\lambda$ with an approximation $\tilde A_\lambda$ can cause transitions out of the instantaneous ground state, with the resulting ground-state infidelity bounded by \cite{finzgar2025counterdiabatic},
\begin{equation}\label{eq:ground_fidelity}
    1-\mathcal F\leq 4\int_0^1d\lambda\sum_{n>0}|\langle\phi_n|(A_\lambda-\tilde A_\lambda)|\phi_0\rangle|^2.
\end{equation}
Since only ground-to-excited-state transitions contribute, the relevant spectral scale is the ground-state gap $\Delta=\min_{\lambda\in[0,1]}\omega_{10}>0$, rather than the minimum spacing $\omega_{\rm min}$ of the full spectrum. 

To obtain a Hermitian AGP, suitable for implementation as a counterdiabatic Hamiltonian on a quantum device~\cite{Claeys19,hegade22,hatomura2026universaldigitizedcounterdiabaticdriving}, polynomial expansions in the Liouvillian are restricted to odd powers.
Ref.~\cite{finzgar2025counterdiabatic} shows that $K=\mathcal O[(B/\Delta)\log(N/\epsilon)]$ odd Chebyshev terms, corresponding to polynomial degree $2K-1$, suffice to achieve a ground-state fidelity $\mathcal F\geq1-\epsilon$, up to logarithmic corrections.
For a local Hamiltonian with extensive bandwidth $B=\mathcal O(N)$, this gives
\begin{equation}\label{eq:global_order}
    K_{\rm global}
    =\mathcal O\left(\frac{N}{\Delta}\log\frac{N}{\epsilon}\right).
\end{equation}
Since $K$ sets the maximum nested-commutator (or Krylov) order $2K-1$, its growth with system size leads to increasingly complex counterdiabatic drives and potentially higher implementation costs. Although suitable gate-based protocols can implement counterdiabatic evolution with polynomial quantum resources for inverse-polynomial gaps~\cite{wan2020fast,vanVreumingen2024,hatomura2026universaldigitizedcounterdiabaticdriving}, these resource requirements can still grow substantially with system size.

In contrast, finite-time filtering and Lieb--Robinson localization restrict each AGP contribution to a region of radius $R=\mathcal O[\Delta^{-1}\log(N/\epsilon)]$, reducing the relevant bandwidth to $B_R=\mathcal O(JR^d)$.
With a filtering time $T=\mathcal O[\Delta^{-1}\log(N/\epsilon)]$, each contribution can be approximated using $K=\mathcal O(B_RT+\log(N/\epsilon))$ odd Chebyshev terms, giving
\begin{equation}\label{eq:local_order}
    K_{\rm local}
    =\mathcal O\left(\frac{1}{\Delta^{d+1}}\log^{d+1}\frac{N}{\epsilon}\right),
\end{equation}
up to additional logarithmic corrections and fixed microscopic scales, as derived in the End Matter.
For a fixed ground-state gap, this reduces the sufficient expansion order from nearly linear to polylogarithmic in system size.

To illustrate the benefit of localization, we compare global and localized Chebyshev approximations for ground-state transport in the non-integrable mixed-field Ising chain~\cite{finzgar2025counterdiabatic},
\begin{equation}
    H(\lambda)=-2J\sum_{j=1}^{L-1}Z_jZ_{j+1}
    +h_z\sum_{j=1}^{L}Z_j
    +h_x(\lambda)\sum_{j=1}^{L}X_j,
\end{equation}
with $J=1$, $h_z=1/\varphi$, and $h_x(\lambda)=1/2+2\lambda$, where $\varphi$ is the golden ratio.
We consider small system sizes $L=6$--$12$, allowing the ground-state infidelities to be evaluated exactly.
For both global and localized expansions, we approximate $1/\omega$ directly, using the minimum ground-state gap $\Delta>0$ of the full Hamiltonian along the transport path and the corresponding global or restricted bandwidth.
As shown in Fig.~\ref{fig:figure1}(a), for $L=12$ localized approximations achieve lower infidelities at substantially smaller expansion orders. Each localized curve is shown up to its minimum infidelity, as higher orders probe timescales beyond those reliably captured within the Lieb--Robinson restricted region of radius $R$. For a target infidelity $\epsilon=10^{-2}$, Fig.~\ref{fig:figure1}(b) shows a nearly linear increase in the required Chebyshev order for the global approximation, compared with a much weaker system-size dependence for the localized expansion. The contrasting bandwidth scaling of the global and restricted Hamiltonians shows that the advantage of localization will become increasingly pronounced with system size.

\emph{Krylov-space complexity--}~
The improvement from finite time filtering and localization can be understood through the suppression of operator growth in Krylov space~\cite{parker19,Takahashi24}. In the Lanczos representation, Heisenberg evolution corresponds to wavepacket propagation along a one-dimensional chain, where the sites $n$ correspond to orthogonal polynomials of the Liouvillian acting on the initial operator $P_n(\mathcal{L_\lambda})O_j$ and the hopping amplitudes are given by the Lanczos (recurrence) coefficients $b_n$ (see End Matter). The reduced polynomial order therefore corresponds to a smaller Krylov subspace needed to approximate the AGP.

\begin{figure}[t]
    \centering
    \includegraphics[width=\columnwidth]{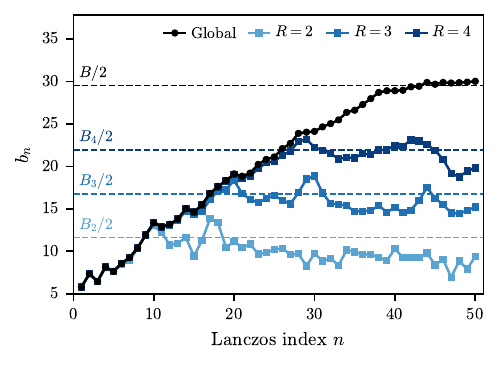}
    \caption{
    Lanczos coefficients $b_n$ for the mixed-field Ising chain with $L=12$ at $\lambda=0.5$, initialized with the local driving term $O_6=2X_6$ entering $\partial_\lambda H$.
    Dashed lines indicate the half-bandwidths $B_R/2$ of the restricted Hamiltonians and $B/2$ of the full Hamiltonian.
    }
    \label{fig:krylov}
\end{figure}

In many-body systems, the Lanczos coefficients typically grow linearly $b_n \approx cn$ (see Fig.~\ref{fig:krylov}), resulting in exponential growth of the Krylov complexity $\left\langle n \right\rangle \sim e^{ct}$, reflecting rapid operator spreading \cite{parker19}. 
However, the Lanczos coefficients are bounded by the Liouvillian norm, $b_n\leq B=\mathcal O(N)$, so this linear growth cannot persist indefinitely.
This implies that the coefficients saturate at $n^\ast=\mathcal{O}(N)$ as shown in the global curve in Fig.~\ref{fig:krylov}.
Within this saturation regime, the wavepacket only spreads ballistically, rather than exponentially, although with an extensive hopping rate. 
The linear scaling of the global expansion order in Eq.~\eqref{eq:global_order} can therefore be understood as localization of the AGP in Krylov space within this saturation regime.

Finite-time filtering alone limits the duration of operator evolution but leaves the Krylov-space dynamics unchanged, and still exhibits exponential growth of the Krylov complexity.
However, the required evolution time $T = \mathcal{O}(\Delta^{-1}\log N)$ allows for the Hamiltonian to be replaced by one with bandwidth $B_R=\mathcal{O}(\log^d N)$ which instead modifies the Lanczos coefficient growth.
As shown in Fig.~\ref{fig:krylov},  the restricted Lanczos coefficients approach smaller plateaus at much earlier Krylov depths of $n_R^\ast=\mathcal{O}(\log^dN)$.
This earlier plateauing limits the regime of rapid operator growth, allowing the AGP to be approximated within a dramatically smaller Krylov subspace and thereby reducing the polynomial order while preserving the same target ground-state fidelity.
Since Chebyshev polynomials correspond to constant Lanczos coefficients $b_n=1/2$ in rescaled units, our convergence bound ignores additional structure in the actual coefficients that could potentially improve the approximation.

\emph{Local observable transport---}~
The accuracy requirements of the AGP approximation can be relaxed considerably when only local observables are of interest.
As detailed in the End Matter, all approximation parameters can then be chosen independently of the system size, consistent with established results on local ground-state transport in gapped systems~\cite{hastings05,osborne2007simulating,bachman12,huang2020computinglocalpropertiestrivial}.
To achieve an accuracy $\epsilon\|Q\|_{\rm op}$ for a local observable $Q$, it is sufficient to retain AGP contributions within a distance
\begin{equation}
    \ell=\mathcal O\left(
    \frac{1}{\Delta^{d+2}}
    \log^{d+2}\frac{1}{\Delta\epsilon}
    \right)
\end{equation}
of its support.
For a fixed spectral gap, the filtering time, polynomial order, and number of spins required for AGP construction and observable transport are all independent of the system size and grow only polylogarithmically with the inverse target accuracy.

Our construction also provides a systematic approach to classical quasi-adiabatic continuation, complementing numerical methods that have proved useful for mapping phase diagrams and probing critical behavior beyond the regime of rigorous guarantees~\cite{kim24,keever24,Medvidovic26}.
Since the evolution is restricted to $\mathcal O((\ell+R)^d)$ spins, even brute-force classical simulation requires resources scaling as $\exp[\mathcal O(\ell^d)]$, which is quasipolynomial in the inverse target accuracy for a fixed spectral gap and independent of the total system size. 
Furthermore, classical state transport does not require a Hermitian generator, allowing the polynomial approximation to include both even and odd orders leading to a quadratic reduction of the needed polynomial order~\cite{finzgar2025counterdiabatic}.

\emph{Discussion and outlook --}~
In summary, we have shown that exploiting spatial locality substantially reduces the resources needed to approximate the AGP. The resulting local AGP expansions remain compatible with existing Floquet-Trotter implementations ~\cite{Claeys19,hatomura2026universaldigitizedcounterdiabaticdriving}.
Although the universal Chebyshev expansion used here establishes rigorous convergence guarantees, the spatial localization of the AGP is independent of the subsequent approximation scheme and can be combined with other existing approaches. Further improvements may be achieved with little additional computational overhead by exploiting spectral structure, either by computing the exact AGP in the restricted Krylov space or a suitably localized variational ansatz. 

Our results improve the efficiency of quantum state preparation through counterdiabatic driving and provide a construction for transporting local observables to explore phase diagrams.
The ability to efficiently approximate the AGP further enables the computation of the quantum metric tensor $g_{ij}(\lambda)=\left\langle \phi_0 | A_i A_j|\phi_0 \right\rangle_c.$
For an initial product state, connected correlations of the spatially truncated AGP contributions vanish at separations beyond $\mathcal O(\ell+R)$.
The remaining correlations can therefore be evaluated on $\mathcal O((\ell+R)^d)$ spins, retaining the quasipolynomial resource scaling of local observable transport.

Knowledge of the quantum geometry enables adaptive integration along Hamiltonian paths~\cite{tomka2016geodesic,kim24} and the detection of phase transitions without prior knowledge of an order parameter~\cite{wang15}.
Near a continuous phase transition, the singular part of the metric density exhibits universal scaling, $g/N\sim|\lambda-\lambda_c|^{\nu d-2}$~\cite{venuti07,polkovnikov05}.
While classical transport becomes increasingly costly as the gap closes, quantum transport remains efficient for algebraically closing gaps~\cite{wan2020fast,vanVreumingen2024}.
On quantum devices, the quantum metric can instead be accessed directly through the nonadiabatic work fluctuations induced by a slow parameter ramp, $\Delta E^2=\sum_{ij}g_{ij}\dot{\lambda}_i \dot{\lambda}_j +\mathcal{O}(\dot{\lambda}^3)$~\cite{funo17,ozawa18,KOLODRUBETZ20171}.
During the finalization of this work, a related approach focusing on certified discovery of quantum phase diagrams was proposed in Ref.~\cite{liu2026certifieddiscoveryquantumphase}.

\emph{Acknowledgements}
A.O. acknowledges support from the Pierre Hohenberg Graduate Scholar Fellowship and computational resources provided by the Flatiron Institute. The Flatiron Institute is a division of the Simons Foundation. D.S. was supported by AFOSR 
under Award No. FA9550-21-1-0236.
\bibliography{references}

\newpage
\appendix
\onecolumngrid
\begin{center}
    \textbf{End Matter}
\end{center}
\twocolumngrid

\emph{Details of ground-state transport---}
Here we derive error bounds for ground-state transport under the three approximations introduced in the main text: finite-time filtering, Lieb--Robinson localization, and Chebyshev expansion.
We assume that the geometrically local Hamiltonian $H(\lambda)$ has a nondegenerate ground state separated from the excited states by a finite gap $\Delta=\min_{\lambda\in[0,1]}\omega_{10}>0$, and that $\partial_\lambda H$ consists of $\mathcal O(N)$ bounded local terms.
Using the infidelity bound in Eq.~\eqref{eq:ground_fidelity}, we determine the number $K$ of odd Chebyshev terms, corresponding to a polynomial degree $2K-1$, sufficient to achieve a target fidelity $\mathcal F\geq 1-\epsilon$.

For the finite-time AGP approximation $A_{\lambda,T}$ defined in Eq.~\eqref{eq:finite_time_agp}, choosing $T=2M/\Delta$ gives
\begin{equation}
    \begin{aligned}
    &\sum_{n>0}
    |\langle\phi_n|
    A_\lambda-A_{\lambda,T}
    |\phi_0\rangle|^2\\
    &=
    \sum_{n>0}
    \frac{|f_T(\omega_{n0})|^2}{\omega_{n0}^2}
    |\langle\phi_n|\partial_\lambda H|\phi_0\rangle|^2\\
    &\leq
    \frac{4^{-M}}{\Delta^2}
    \operatorname{Var}_{\phi_0}(\partial_\lambda H),
    \end{aligned}
    \label{eq:filtering_error}
\end{equation}
where we used $|f_T(\omega)|\leq 2^{-M}$ for $|\omega|\geq\Delta$.
Thus, the filtering error depends only on the ground-state gap, without requiring a minimum spacing between excited states.

For the localized AGP defined in Eq.~\eqref{eq:localized_agp}, the Lieb--Robinson bound gives
\begin{equation}
    \|A_{j,T}-A_{j,T,R}\|_{\rm op}
    \leq CT e^{-\mu(R-vT)},
\end{equation}
where $C$, $\mu$, and $v$ are independent of system size.
Summing over the $\mathcal O(N)$ local contributions yields
\begin{equation}
    \|A_{\lambda,T}-A_{\lambda,T,R}\|_{\rm op}
    \leq CNT e^{-\mu(R-vT)}.
    \label{eq:localization_error}
\end{equation}

Following the Chebyshev approximation approach of Ref.~\cite{finzgar2025counterdiabatic}, we approximate each localized AGP contribution by a polynomial in the restricted Liouvillian $\mathcal L_{R,j}=[H_{R,j},\cdot]$.
The restricted Hamiltonian has bandwidth $B_R=\mathcal O(JR^d)$, determined by the localization radius rather than the total system size.
Since the filtered function $g_T(\omega)$ is odd, its Chebyshev expansion contains only odd powers,
\begin{equation}
    g_T(\mathcal L_{R,j})
    =\sum_{k=1}^{\infty}a_{2k-1}
    T_{2k-1}\left(\frac{\mathcal L_{R,j}}{B_R}\right),
\end{equation}
where $T_k$ denotes the Chebyshev polynomial of the first kind.
Truncating after $K$ terms defines the approximate AGP
\begin{equation}
    A_{\lambda,T,R,K}
    =-i\sum_j\sum_{k=1}^{K}a_{2k-1}
    T_{2k-1}\left(\frac{\mathcal L_{R,j}}{B_R}\right)O_j.
\end{equation}
Since $g_T(z)$ is entire and satisfies $|g_T(z)|\leq T e^{T|\operatorname{Im}z|}$, standard Chebyshev approximation bounds~\cite{trefethen2020approximation} give
\begin{equation}
    |a_n|\leq 2T
    \exp\left[
    \frac{B_RT}{2}(\rho-\rho^{-1})
    \right]\rho^{-n},
    \label{eq:chebyshev_coeff}
\end{equation}
where $\rho>1$ parametrizes a Bernstein ellipse.
The Chebyshev recurrence $T_{n+1}(x)=2xT_n(x)-T_{n-1}(x)$, together with $\|\mathcal L_{R,j}(O)\|_{\rm op}\leq B_R\|O\|_{\rm op}$, implies
\begin{equation}
    \left\|
    T_n\left(\frac{\mathcal L_{R,j}}{B_R}\right)O_j
    \right\|_{\rm op}
    \leq(1+\sqrt{2})^n\|O_j\|_{\rm op}.
\end{equation}
Choosing any fixed $\rho>1+\sqrt{2}$ and bounding the truncation error by the remaining Chebyshev coefficients gives
\begin{equation}
    \|A_{\lambda,T,R}-A_{\lambda,T,R,K}\|_{\rm op}
    \leq CNT e^{c_1B_RT-c_2K},
    \label{eq:chebyshev_error}
\end{equation}
where $C,c_1,c_2>0$ are independent of system size.

Combining Eqs.~\eqref{eq:filtering_error}, \eqref{eq:localization_error}, and \eqref{eq:chebyshev_error} with the ground-state infidelity bound, a sufficient condition for $\mathcal F\geq1-\epsilon$ is
\begin{equation}
    \begin{aligned}
    \frac{2^{-M}\sqrt{\mathcal V}}{\Delta}
    &\lesssim\sqrt{\epsilon},\\
    NT e^{-\mu(R-vT)}
    &\lesssim\sqrt{\epsilon},\\
    NT e^{c_1B_RT-c_2K}
    &\lesssim\sqrt{\epsilon},
    \end{aligned}
    \label{eq:app_error_conditions}
\end{equation}
where
$\mathcal V=\int_0^1d\lambda\,
\operatorname{Var}_{\phi_0}(\partial_\lambda H)$.
For bounded local interactions, $\mathcal V=\mathcal O(N^2)$ provides a sufficient upper bound.
For a ground-state gap bounded below independently of system size, the first two conditions give
\begin{equation}
    T=\mathcal O\left(
    \frac1\Delta\log\frac{N}{\epsilon}
    \right),
    \qquad
    R=\mathcal O\left(
    \frac1\Delta\log\frac{N}{\epsilon}
    \right),
\end{equation}
where fixed microscopic scales and additional gap-dependent logarithmic factors are suppressed.
Using $B_R=\mathcal O(JR^d)$, the Chebyshev truncation error is controlled by an expansion order
\begin{equation}
    \begin{aligned}
    K_{\rm local}
    &=\mathcal O\left(
    B_RT+\log\frac{N}{\epsilon}
    \right)\\
    &=\mathcal O\left(
    \frac1{\Delta^{d+1}}
    \log^{d+1}\frac{N}{\epsilon}
    \right).
    \end{aligned}
\end{equation}
This establishes Eq.~\eqref{eq:local_order} in the main text, showing that the sufficient expansion order grows only polylogarithmically with system size for a fixed ground-state gap.

\emph{Operator growth in Krylov space---}~
Since the AGP is approximated by a polynomial in the Liouvillian, its expansion can be understood in terms of the Krylov space generated by repeated applications of $\mathcal L$ to a local driving term $O_j$,
\begin{equation}
    \mathcal K_m(\mathcal L,O_j)
    =\operatorname{span}\{O_j,\mathcal LO_j,\ldots,\mathcal L^{m-1}O_j\}.
\end{equation}
An expansion containing $K$ odd Chebyshev terms has degree $2K-1$ and therefore belongs to $\mathcal K_{2K}(\mathcal L,O_j)$.
An orthonormal Krylov basis can be constructed using the Lanczos algorithm~\cite{muck2022krylov},
\begin{equation}
    O_n=P_n(\mathcal L)O_0, \qquad O_0=O/\|O\|,
\end{equation}
where $P_n$ are polynomials orthonormal with respect to the spectral measure of $O$. Using the normalized Hilbert--Schmidt inner product $(A|B)=\operatorname{Tr}(A^\dagger B)/D$, with $\|O\|^2=(O|O)$, their orthogonality is expressed as
\begin{equation}
    \int d\mu_O(\omega)\,P_m(\omega)P_n(\omega) =(O_m|O_n)=\delta_{mn},
\end{equation}
where $\int d\mu_O(\omega)f(\omega)=(O_0|f(\mathcal L)O_0)$. For Hermitian $O$, the polynomials satisfy the three-term recurrence
\begin{equation}
    \omega P_n(\omega)=b_{n+1}P_{n+1}(\omega)+b_nP_{n-1}(\omega),
\end{equation}
with $P_0(\omega)=1$, $P_{-1}(\omega)=0$, and $b_0=0$. The Lanczos coefficients $b_n$ are the off-diagonal elements of the Liouvillian in this basis, and govern the growth of operators in Krylov space.

The Heisenberg-evolved operator can be expanded in this basis as
\begin{equation}
    O_j(t)=e^{i\mathcal Lt}O_j
    =\|O_j\|\sum_{n\geq0}i^n\phi_n(t)O_n,
\end{equation}
where the amplitudes satisfy
\begin{equation}
    \dot\phi_n(t)
    =b_n\phi_{n-1}(t)-b_{n+1}\phi_{n+1}(t),
    \qquad \phi_n(0)=\delta_{n0}.
\end{equation}
Thus, Heisenberg evolution corresponds to wavepacket propagation along a one-dimensional chain, with hopping amplitudes given by the Lanczos coefficients $b_n$.
Since Heisenberg evolution preserves the Hilbert--Schmidt norm, the amplitudes satisfy $\sum_{n\geq0}|\phi_n(t)|^2=1$.
The average position of this wavepacket defines the Krylov complexity~\cite{parker19},
\begin{equation}
    \langle n\rangle(t)
    =\sum_{n\geq0}n|\phi_n(t)|^2,
\end{equation}
which measures the average depth reached by an operator in Krylov space.

For linearly growing Lanczos coefficients, $b_n\sim\alpha n$, the Krylov complexity grows exponentially, $\langle n\rangle\sim e^{2\alpha t}$~\cite{parker19}. However, the Lanczos coefficients are bounded by the Liouvillian norm,
\begin{equation}
    b_n\leq\|\mathcal L\| =B\leq2\|H\|_{\rm op},
\end{equation}
where $B$ is the Hamiltonian bandwidth. Consequently, linear Lanczos growth cannot persist indefinitely in a finite system and $\langle n\rangle(t)\leq2B|t|$. Restricting the Hamiltonian to a finite region replaces $B$ by the smaller bandwidth $B_R$, limiting the rate of Krylov spreading.

\emph{Local observable transport---}~
We consider the transport of a local observable $Q$ along a gapped Hamiltonian path.
Let $A_\lambda=\sum_j A_{j,\lambda}$ denote the exact AGP, decomposed according to the local driving terms $O_j$ in $\partial_\lambda H$.
We exploit locality by retaining only the approximate AGP contributions in the vicinity of $Q$,
\begin{equation}
    \widetilde A_\lambda
    =\sum_{j\in\Lambda_\ell}\widetilde A_{j,\lambda},
\end{equation}
where $\Lambda_\ell$ contains the driving terms within distance $\ell$ of the support of $Q$.
Each approximate AGP contribution has norm $\mathcal O(T)$, support radius $R$, and overlaps with at most $\mathcal O(R^d)$ other contributions.
The corresponding Lieb--Robinson velocity bound scales as
\begin{equation}
    v_{\rm AGP}=\mathcal O(TR^{d+1}).
\end{equation}
To capture the spreading of $Q$ under the AGP-generated evolution, we choose a cutoff $\ell\gtrsim TR^{d+1}$ extending beyond the effective Lieb--Robinson light cone.

We define the backward-evolved observable
\begin{equation}
    Q_\lambda=U(1,\lambda)^\dagger Q U(1,\lambda),
    \qquad
    U(1,\lambda)=\mathcal P
    e^{-i\int_\lambda^1 d\lambda'\,\widetilde A_{\lambda'}}.
\end{equation}
Our approximation to the final ground-state expectation value $\langle Q_1\rangle_1$ is then given by $\langle Q_0\rangle_0$, where $\langle\cdot\rangle_\lambda$ denotes the expectation value in $|\phi_0(\lambda)\rangle$.
To bound the difference, we use
\begin{equation}
    \frac{d}{d\lambda}\langle Q_\lambda\rangle_\lambda
    =i\langle[A_\lambda-\widetilde A_\lambda,Q_\lambda]\rangle_\lambda.
\end{equation}
Integrating along the path and decomposing the AGP into its local contributions gives
\begin{equation}
    \begin{aligned}
    \epsilon_Q
    &\equiv|\langle Q_1\rangle_1-\langle Q_0\rangle_0|\\
    &\leq\int_0^1d\lambda
    \sum_{j\in\Lambda_\ell}
    |\langle[A_j-\widetilde A_j,Q_\lambda]\rangle_\lambda|\\
    &\quad+\int_0^1d\lambda
    \sum_{j\notin\Lambda_\ell}
    |\langle[A_j,Q_\lambda]\rangle_\lambda|\\
    &\equiv\epsilon_{\rm in}+\epsilon_{\rm out}.
    \end{aligned}
\end{equation}
Here, $\epsilon_{\rm in}$ accounts for the approximation of the retained AGP contributions, while $\epsilon_{\rm out}$ accounts for their spatial truncation.

Using the approximate AGP construction from above, we can bound $\epsilon_{\rm in}$ using Eqs.~\eqref{eq:filtering_error}, \eqref{eq:localization_error}, and \eqref{eq:chebyshev_error}.
Defining the maximum ground-state error per contribution as
\begin{equation}
    \begin{aligned}
    \eta(T,R,K)
    &=\sup_{\lambda,j\in\Lambda_\ell}
    \|(A_{j,\lambda}-\widetilde A_{j,\lambda})
    |\phi_0(\lambda)\rangle\|\\
    &\lesssim
    \frac{2^{-\Delta T/2}}{\Delta}
    +Te^{-\mu(R-vT)}
    +Te^{c_1B_RT-c_2K},
    \end{aligned}
\end{equation}
we obtain
\begin{equation}
    \epsilon_{\rm in}
    \leq 2|\Lambda_\ell|\eta\|Q\|_{\rm op}
    \lesssim\ell^d\eta\|Q\|_{\rm op},
\end{equation}
where we used $|\Lambda_\ell|=\mathcal O(\ell^d)$ for a local observable in $d$ dimensions.

Since the exact AGP contributions need not be local, we introduce an auxiliary finite-time approximation $A_{j,T_j}$ for each omitted contribution to bound $\epsilon_{\rm out}$.
Adding and subtracting this approximation gives
\begin{equation}
    \begin{aligned}
    |\langle[A_j,Q_\lambda]\rangle_\lambda|
    &\leq 2\|Q\|_{\rm op}
    \|(A_j-A_{j,T_j})|\phi_0(\lambda)\rangle\|\\
    &\quad+\|[A_{j,T_j},Q_\lambda]\|_{\rm op}.
    \end{aligned}
\end{equation}
While increasing $T_j$ suppresses the filtering error in the first term, it also increases the spatial extent of $A_{j,T_j}$.
Following Ref.~\cite{hastings05}, we therefore choose the filtering time to increase with the distance $D_j=\operatorname{dist}(\operatorname{supp}O_j,\operatorname{supp}Q)$.
This allows for increasingly accurate approximations of distant AGP contributions while keeping their effective Lieb--Robinson light cones separated from that of $Q_\lambda$.
Specifically, we choose
\begin{equation}
    M_j=\left\lfloor\frac{\Delta D_j}{8v}\right\rfloor,
    \qquad
    T_j=\frac{2M_j}{\Delta},
\end{equation}
so that $vT_j\leq D_j/4$.
For sufficiently large $D_j$, the per-term filtering bound gives
\begin{equation}
    \|(A_j-A_{j,T_j})|\phi_0(\lambda)\rangle\|
    \lesssim\frac{1}{\Delta}e^{-c\Delta D_j/v},
\end{equation}
where $c>0$ is independent of system size.

We now show that the remaining commutator is small because the two operators remain approximately separated under their respective evolutions.
The finite-time AGP $A_{j,T_j}$ is constructed from Heisenberg-evolved driving terms $O_j(s)$ with $|s|\leq T_j$.
Its commutator can therefore be bounded by
\begin{equation}
    \|[A_{j,T_j},Q_\lambda]\|_{\rm op}
    \leq T_j\sup_{|s|\leq T_j}
    \|[O_j(s),Q_\lambda]\|_{\rm op}.
\end{equation}

Since $vT_j\leq D_j/4$, the Lieb--Robinson bound for $H(\lambda)$ allows $O_j(s)$ to be approximated by an operator supported within radius $D_j/2$ of its original support, with error
\begin{equation}
    \|O_j(s)-O_j^{(D_j/2)}(s)\|_{\rm op}
    \lesssim e^{-\mu D_j/4}.
\end{equation}
Similarly, using the Lieb--Robinson velocity $v_{\rm AGP}=\mathcal O(TR^{d+1})$ and the interaction range $\mathcal O(R)$ of the approximate AGP, we can localize $Q_\lambda$ within a radius $D_j/4$ with error
\begin{equation}
    \|Q_\lambda-Q_\lambda^{(D_j/4)}\|_{\rm op}
    \lesssim \|Q\|_{\rm op}
    e^{-\kappa D_j/R+\gamma TR^d},
\end{equation}
where $\kappa,\gamma>0$ are independent of $T$, $R$, and the system size.

The two localized operators have disjoint supports and therefore commute exactly.
The commutator of the original operators is consequently bounded by their localization errors.
Combining these estimates with the filtering error gives
\begin{equation}
    \begin{aligned}
    |\langle[A_j,Q_\lambda]\rangle_\lambda|
    \lesssim\|Q\|_{\rm op}\bigg[
    &\Delta^{-1}e^{-c\Delta D_j/v}\\
    &+T_j e^{-\mu D_j/4}\\
    &+T_j e^{-\kappa D_j/R+\gamma TR^d}
    \bigg].
    \end{aligned}
\end{equation}
Summing over the $\mathcal O(D^{d-1})$ driving terms at distance $D$ and using $T_j\leq D_j/(4v)$ gives
\begin{equation}
    \begin{aligned}
    \epsilon_{\rm out}
    \lesssim\|Q\|_{\rm op}
    \sum_{D>\ell}D^{d-1}\bigg[
    &\Delta^{-1}e^{-c\Delta D/v}\\
    &+\frac{D}{v}e^{-\mu D/4}\\
    &+\frac{D}{v}e^{-\kappa D/R+\gamma TR^d}
    \bigg].
    \end{aligned}
\end{equation}
The exponential decay with distance dominates the polynomial growth in the number of driving terms, allowing $\epsilon_{\rm out}$ to be made arbitrarily small independently of the system size.

We now choose the approximation parameters such that $\epsilon_{\rm in},\epsilon_{\rm out}\leq\epsilon\|Q\|_{\rm op}/2$.
Evaluating the exponentially decaying tails gives a sufficient cutoff
\begin{equation}
    \ell\gtrsim TR^{d+1}
    +R\log\frac{R^{d+1}}{\epsilon}
    +\frac{v}{\Delta}\log\frac{1}{\Delta^{d+1}\epsilon},
\end{equation}
where fixed microscopic scales are suppressed.
For the retained contributions, we require
\begin{equation}
    \eta(T,R,K)\lesssim\frac{\epsilon}{\ell^d}.
\end{equation}
Using the error bounds derived above and $B_R=\mathcal O(R^d)$, both conditions can be satisfied self-consistently with
\begin{equation}
    \begin{aligned}
    T,R&=\mathcal O\left(
    \frac1\Delta\log\frac1{\Delta\epsilon}
    \right),\\
    K&=\mathcal O\left(
    \frac1{\Delta^{d+1}}
    \log^{d+1}\frac1{\Delta\epsilon}
    \right),\\
    \ell&=\mathcal O\left(
    \frac1{\Delta^{d+2}}
    \log^{d+2}\frac1{\Delta\epsilon}
    \right).
    \end{aligned}
\end{equation}
Only $\mathcal O(\ell^d)$ approximate AGP contributions need to be constructed, each requiring a region of $\mathcal O(R^d)$ spins.
The resulting AGP evolution is then restricted to only $\mathcal O((\ell+R)^d)$ spins.
For a fixed spectral gap, both the number of retained contributions and the size of the regions required for their construction and subsequent evolution grow only polylogarithmically with the inverse target accuracy, independently of the total system size $N$.

\end{document}